# The Mass of the Milky Way Galaxy

Christopher S. Kochanek

Harvard–Smithsonian Center for Astrophysics MS-51, 60 Garden St., Cambridge, MA 02138

I: ckochanek@cfa.harvard.edu

## ABSTRACT

We use the Jaffe model as a global mass distribution for the Galaxy and determine the circular velocity $v_c$ and the Jaffe radius $r_j$ using the satellites of the Galaxy, estimates of the local escape velocity of stars, the constraints imposed by the known rotation curve of the disk, and the Local Group timing model. The models include the systematic uncertainties in the isotropy of the satellite orbits, the form of the stellar distribution function near the escape velocity, and the ellipticity of the M31/Galaxy orbit. If we include the Local Group timing constraint, then Leo I is bound, $v_c = 230 \pm 30$ km s$^{-1}$, and $r_j = 180$ kpc (110 kpc $\lesssim r_j \lesssim$ 300 kpc) at 90% confidence. The satellite orbits are nearly isotropic with $\beta = 1 - \sigma_\theta^2/\sigma_r^2 = 0.07$ ($-0.7 \lesssim \beta \lesssim 0.6$) and the stellar distribution function near the escape velocity is $f(\epsilon) \propto \epsilon^k$ with $k_r = 3.7$ ($0.8 \lesssim k_r \lesssim 7.6$) where $k_r = k + 5/2$. While not an accurate measurement of $k$, it is consistent with models of violent relaxation ($k = 3/2$). The mass inside 50 kpc is $(5.4 \pm 1.3) \times 10^{11} M_\odot$. Higher mass models require that M31 is on its second orbit and that the halo is larger than the classical tidal limit of the binary. Such models must have a significant fraction of the Local Group mass in an extended Local Group halo. Lower mass models require that both M31 and Leo I are unbound, but there is no plausible mechanism to produce the observed deviations of M31 and Leo I from their expected velocities in an unbound system. If we do not use the Local Group timing model, the median mass of the Galaxy *increases* significantly, and the error bars broaden. Using only the satellite, escape velocity, and disk rotation curve constraints, the median mass interior to 50 kpc is 4.3 (5.4) and the 90% confidence interval is 3.3 to 6.1 (4.2 to 6.8) without (with) Leo I in units of $10^{11} M_\odot$. The lower bound without Leo I is 65% of the mass expected for a continuation of a flat rotation curve.

*Subject headings:* galactic structure – galactic dynamics – dark matter – galactic kinematics – galactic halo – Local Group

# 1 INTRODUCTION

In our own Galaxy the evidence for a large dark matter halo is generally weaker than for external spiral galaxies with extended rotation curves (see Fich & Tremaine 1991). There are three dynamical constraints on our halo: the orbits of satellites, the local escape velocity, and Local Group timing. For each of these constraints we must assume that it applies at all – the satellites, stars, or M31 must be bound to the Galaxy. Each constraint also has a systematic uncertainty: the isotropy for the satellite orbits, the shape of the distribution function near the escape velocity for the stars, and the ellipticity of the orbit and the age of the universe for Local Group timing. Analyses of the halo almost always apply one of these limits in isolation, and the inferences about the halo are strongly limited by the systematic problems of the constraint. Yet models that assume the satellites are bound almost always give the halo a structure that leads to the stars and M31 being bound (and *vice versa*) so we should really examine the properties of a halo consistent with all three constraints. There are also two boundary conditions on the models. At small radii the mass of the model must be consistent with the known properties of the rotation curve of the disk, and at large radii we should consider the effects of M31 on tidally limiting the size of any halo bound solely to the Galaxy.

In this paper we try to build a self-consistent mass distribution for the Galaxy between the outer edge of the disk and M31 including the primary systematic uncertainties on each of the constraints as we proceed. The kinematics of the disk inside 20 kpc are consistent with a flat or slightly rising rotation curve (Fich & Tremaine 1991), and a self-consistent model must be constrained to merge with the known properties of the disk. Any spherical density distribution must approach $\rho = \Theta_0^2/4\pi G r^2$ where $\Theta_0$ is the circular velocity in the disk near the solar radius $R_0$ for $r \lesssim 2R_0$. In general we normalize the rotation velocity and the solar radius to the 1985 IAU values (Swings 1985) of $\Theta_0 = 220$ km s$^{-1}$ and $R_0 = 8.5$ kpc, although recent estimates of $R_0$ are slightly lower (Reid 1993). The fundamental question about the mass distribution of the Galaxy is whether there is a cutoff in the density distribution. The simplest model that can be consistent with the kinematics of the disk and has a cutoff is the Jaffe (1983) model, where

$$\rho = \frac{v_c^2}{4\pi G r^2} \frac{r_j^2}{(r+r_j)^2}. \qquad (1)$$

For $r \lesssim r_j$ the model produces a flat rotation curve and for $r \gtrsim r_j$ the density distribution cuts off as $\rho \propto r^{-4}$. No point mass or power law model can mimic globally a density distribution with a break.

The highest velocity stars near the sun (eg. Carney & Latham 1987, Carney et al. 1994) estimate the local escape velocity $v_e$ of the Galaxy, and the most recent analysis yields 450 km s$^{-1} < v_e < 650$ km s$^{-1}$ (Leonard & Tremaine 1990) where the dominant source of the uncertainty is the shape of the distribution function near the escape velocity. The escape velocity in the Jaffe model is $v_e^2 = 2v_c^2 \ln(1 + r_j/r)$ and for $v_c = \Theta_0 = 220$ km s$^{-1}$ the range for the escape velocity corresponds to 44 kpc $< r_j <$ 660 kpc. The highest velocity star in the sample (assumed to be bound) provides the lower limit on the range, while the upper limit depends on the shape of the distribution function of stellar energies near the escape velocity or the slope of the stellar density at large radii. For a sharp density cutoff there are few stars near the escape velocity, so the escape velocity is significantly larger than the largest observed stellar velocity. Leonard & Tremaine (1990) found that the stellar velocity data alone could not constrain this systematic uncertainty.

There are 26 satellites with measured radial velocities outside the visible disk ($r \gtrsim 20$ kpc), and 15 between the LMC and 250 kpc. There are many models of the orbital dynamics of the LMC/SMC pair and the Magellanic stream (eg. Lynden-Bell & Lin 1977, Davies & Wright 1977, Murai & Fujimoto 1980, Lin & Lynden-Bell 1982, Lin et al. 1995). The results, although largely qualitative, suggest that the halo is larger than 70 kpc and that the circular velocity at the radius of the LMC is of order 240 km s$^{-1}$. The satellites can also be analyzed as an equilibrium dynamical system. The earliest approaches used either the Jeans equations (Hartwick & Sargent 1978) or the phase-averaged properties of elliptical orbits about a point mass (Lynden-Bell, Cannon, & Godwin 1983, Peterson 1985, Olszewski, Peterson, & Aaronson 1986), but these estimates suffered both from systematic problems in the early analysis procedures and large uncertainties in many of the velocity determinations. Little & Tremaine (1987, LT) made an important improvement in the analysis techniques by using the distribution function of the satellites and a Bayesian statistical approach. By using the distribution function LT ensured that the dynamical models were internally self-consistent, and the Bayesian statistical model gives well defined probability distributions for the mass of the Galaxy. Zaritsky et al. (1989, ZI) repeated the LT analysis using improved velocity estimates. The LT and ZI analysis narrowed the primary uncertainties in the analysis to the isotropy of the orbits and the treatment of Leo I, the most distant high velocity satellite. ZI found that the median mass of the Galaxy modeled as a point mass is 2.7 (4.6) for radial (isotropic) orbits without Leo I, and 9.3 (12.5) with Leo I in units of $10^{11} M_\odot$. Kulessa & Lynden-Bell (1992, KL) found similar mass estimates. KL also made the first attempt to determine the isotropy of the orbits and the density power law ($\rho \propto r^{-\gamma}$) from the satellite orbits. The preferred models had $\gamma \sim 2.4$ and the orbits were more isotropic than radial.

If the Galaxy and M31 are a bound, isolated, binary galaxy system, then the Local Group timing argument (Kahn & Woltjer 1959) yields an estimate of the mass of the Local Group subject to the constraints on the observed separation, radial velocity, and the age of the universe. Einasto & Lynden-Bell (1982) generalized the timing model to include estimates of the tangential motions of M31. Raychaudhury & Lynden-Bell (1989) and Peebles et al. (1989) examined the effects of the collapse phase of evolution and perturbations by other Local Group members, with the conclusion that the mass estimates found from the simple Local Group timing estimates were little affected by the complications. It is possible to find models permitting complicated collisional interactions in the Local Group (eg. Byrd et al. 1994) but there is little evidence that such complicated behavior is required and even then deviations from the simple timing model are small. The timing models do not require that the mass of the Local Group is contained in the halos of M31 and the Galaxy rather than in a larger Group halo, but if we divide the timing mass between the Galaxy and M31, the halo must extend to $r_j = 100$-$200$ kpc.

In this paper we try to build a consistent model based on the Jaffe mass distribution for all these constraints simultaneously. The model nominally covers the entire Galaxy, and we gloss over the distinctions between the disk and the halo in the inner regions.[1] The model is a smooth analogue of the "minimal halo" model used by Fich & Tremaine (1991), and in the limit of a large Jaffe radius it is a reasonable approximation to the spherical infall model ($\rho \propto r^{-2.25}$, see Bertschinger 1985 for example), since the differences between $r^{-2}$ and $r^{-2.25}$ develop slowly. Pure power law models or point mass models are inevitably inconsistent either with the structure of the inner regions of the Galaxy or with the outer regions, and a joint analysis of the satellites, the local escape velocity, the disk, and Local Group timing is not possible using the LT, ZI or KL models for the mass distribution.

In §2 we reexamine the satellites of the Galaxy. Since the ZI and KL analyses, the first proper motions for the outer satellites were measured. The highest accuracy measurements are for the LMC (Jones et al. 1994) and Sculptor (Schweitzer & Cudworth 1995), there are lower accuracy measurements for Pal 3 (Cudworth 1993), and rough estimates for Draco & Ursa Minor (Scholz & Irwin 1994). We use a Bayesian analysis similar to LT, generalized to include estimating the isotropy of the satellite orbits using procedures similar to KL and Merritt & Saha (1993). In this section we focus on interpreting the satellite data so we confine the analysis to a Galaxy normalized by $v_c = \Theta_0$ and estimating the Jaffe radius $r_j$ and the isotropy $\beta$ from the data. In §2.1 we repeat the standard analysis of the radial velocities of the satellites, and in §2.2 we examine the effects of including the proper motion measurements. In §3 we try to estimate the global structure of the Galaxy by fitting both $v_c$ and $r_j$. We start using only the satellite data, and then add the local escape velocity, the constraints from the known rotation curve of the disk, the effects of truncating the halo at the "tidal radius" created by M31, and the Local Group timing model. The advantage of analyzing all the constraints simultaneously is that some of the systematic uncertainties in any one approach are reduced by the constraints from the others. Finally, in §4 we summarize the conclusions.

## 2  The Satellites of the Galaxy

The potential of the Jaffe (1983) model is $\phi = -v_c^2 \ln(1 + r_j/r)$ where $r_j$ is the scale radius at which the density shifts from a singular isothermal density profile ($\rho \propto r^{-2}$) to a steeply declining profile ($\rho \propto r^{-4}$). The total mass inside radius $r$ is $(v_c^2 r_j/G)(1 + r_j/r)^{-1}$. We model the number density distribution of the satellites by a second Jaffe (1983) model $\nu_{sat} \propto r^{-2}(r_s+r)^{-2}$. Several studies (Lake & Tremaine 1980, Lorrimer et al. 1993, Zaritsky et al. 1993) found that the number density of satellites declines with radius as $r^{-1.8\pm0.2}$, closely matching the inner portion of the Jaffe (1983) distribution. KL estimated that the density of satellites of our galaxies declines more steeply ($\propto r^{-3.85}$), but the number distribution of satellites in our Galaxy depends strongly on ill-defined selection effects and magnitude limits that are not included in the model and are likely to make the observed distribution steeper than the true distribution. While it is likely that the number distribution observed in other galaxies is more reliable, we included a scale radius in the satellite density so we can examine its effects. In the limit that the Jaffe scale radius of the satellites is small ($r_s \lesssim R_0$) the density profile of the Jaffe model will resemble that used by KL.

---

[1]This is a reasonable assumption since the only disk model producing a flat rotation curve (the Mestel (1963) disk) has the same mass interior to a given radius as the spherical model producing a flat rotation curve (the singular isothermal sphere). For an exponential disk, the maximum error in using $GM(r)/r$ to estimate the circular velocity is 15%. Compared to the uncertainties in the mass interior to a given radius, the effects of asphericity are negligible.

For simplicity we assume a distribution function of the form $f(\epsilon, \ell) = f(\epsilon)\ell^{-2\beta}$ where $\epsilon = \psi - v^2/2$ is the binding energy per unit mass, $\ell$ is the angular momentum per unit mass, $\psi = -\phi$, and $0 < \epsilon < \psi$. This distribution function produces spherical models with equal velocity dispersions in the orthogonal angular directions ($\sigma_\theta^2 = \sigma_\phi^2$), and a constant orbital anisotropy $\beta = 1 - \sigma_\theta^2/\sigma_r^2$ (eg. Binney & Tremaine 1987). While it has many shortcomings as a global model of the satellite distribution function, it is probably an adequate model for the limited radial range we study, and it has the virtue of containing previous models of the galactic halo in the isotropic and radial limits. Given the bound satellite density distribution $\nu_{sat}(r)$ and the halo potential function $\phi(r)$ the energy dependent part of the distribution function is

$$f(\epsilon) = \frac{2^{\beta-3/2}}{\pi^{3/2}\Gamma[m-1/2-\beta]\Gamma[1-\beta]} \frac{d}{d\epsilon} \int_0^\epsilon d\psi \frac{d^m r^{2\beta} \nu_{sat}}{d\psi^m} (\epsilon - \psi)^{\beta-3/2+m} \quad (2)$$

where $\Gamma[x]$ is the Gamma function and $m$ is an integer. The standard form for this integral uses $m = 1$ (e.g. Dejonghe 1986), but by choosing the value of $m$ we keep the integrand finite and easily integrable with numerical techniques. The distribution function is positive definite for all combinations of scale radii in the two Jaffe models for the mass distribution and satellite number density. The probability that a satellite at radius $r$ has radial velocity $v_r$ is

$$P(v_r|r_j, \beta) = \nu_{sat}^{-1} \int d^3v f(\epsilon)\ell^{-2\beta}\delta(v_r - v_r) = \frac{1}{\sqrt{2\pi}\nu_{sat}r^{2\beta}} \int_0^{\epsilon_m} d\psi \frac{dr^{2\beta}\nu_{sat}}{d\psi}(\epsilon_m - \psi)^{-1/2} \quad (3)$$

if $\epsilon_m = \psi - v_r^2/2 > 0$ and zero otherwise. The notation $P(v_r|\beta, r_j)$ means the probability of finding $v_r$ given the parameters of the model ($\beta$ and $r_j$). In the cases where we have proper motions for the satellites and can infer the total tangential velocity $v_t$, we find $P(v_r, v_t|\beta, r_j) = \nu_{sat}^{-1} f(\epsilon)\ell^{-2\beta}$ if $\psi - (v_r^2 + v_t^2)/2 > 0$, and zero otherwise. For two Jaffe models we can easily determine the radius as a function of the potential, and both the probability of a given radial velocity or the probability of a radial and a tangential velocity can be reduced to a single, convergent, one-dimensional numerical integral. The model depends on four parameters: the inner circular velocity $v_c$, the Jaffe scale radius of the mass density $r_j$, the average isotropy parameter $\beta$, and the Jaffe scale radius of the satellites $r_s$. In this section we hold $v_c = \Theta_0 = 220$ km s$^{-1}$ fixed and explore the relation between the isotropy $\beta$ and the Jaffe radius $r_j$ for several different models of the satellite Jaffe radius $r_s$.

We examined three models for the satellite cutoff radius. In the first model the cutoff radius is a constant, and we examined $r_s = 50$, 100, 250, and 500 kpc. Here the satellite distribution is not modified by changes in the bound matter distribution. In the second model the matter and satellite distributions are identical in shape and $r_s = r_j$. In the last model we assume that the scale of the bound satellite distribution is the current turn-around radius for a top-hat perturbation with a mass equal to the virialized mass of the Galaxy $M = v_c^2 r_j/G$. In the White & Zaritsky (1992) halo model the turn-around radius today is

$$r_t = 635 g \left(\frac{v_c}{200 \text{ km s}^{-1}}\right)^{2/3} \left(\frac{r_j}{100 \text{ kpc}}\right)^{1/3} h_{75}^{-2/3} \text{kpc} \qquad (4)$$

where $H_0 = 75 h_{75}$ km s$^{-1}$ Mpc$^{-1}$ is the Hubble constant. The factor $1 \leq g \leq (3/2)^{2/3}$ is a function of the cosmological model, with the lower limit for $\Omega = 1$ and the upper limit for $\Omega = 0$. We generically used the estimate of $r_t$ with $g = 1$. In practice, the mass inside the turn-around radius is larger than the current virialized mass of the Galaxy, so by setting the mass interior to $r_t$ to be the Galaxy mass we underestimate the true turn-around radius if matter is still settling onto the Galaxy. On the other hand, we are only setting the scale of the turn over in the bound satellite distribution to be $r_t$, so we are using a density distribution extending beyond $r_t$. These cases are designed to illustrate the effects of the model for the satellite distribution on the estimates of the mass of the Galaxy. For the Milky Way, the turn-around radius is larger than half the distance to M31 if $r_j \gtrsim 16 h_{75}^2$ kpc, so the model is of dubious applicability for our Galaxy. We will treat the model with constant $r_s = 250$ kpc as our standard model, although we present the results for the other model assumptions.

Following LT we use a Bayesian statistical approach to study the satellites. For each of $N$ satellite galaxies we measure the line of sight velocity $v_{losi}$ and distance from the sun $r_{\odot i}$ and then estimate the galactocentric radial velocity $v_{ri}$ and radius $r_i$. Then we use equation (2) to compute the probability of finding the observed radial velocity at the observed radius given the model parameters $P(v_{ri}|r_j,\beta)$. We then use Bayes theorem to compute the probability of the parameters given the data and prior information $I$, $P(r_j,\beta|v_{ri},I)$,

$$P(r_j,\beta|v_{ri},I) = \frac{P(r_j)P(\beta)\Pi_{i=1}^N P(v_{ri}|r_j,\beta)}{\int dr_j d\beta P(r_j) P(\beta|r_j) \Pi_{i=1}^N P(v_{ri}|r_j,\beta)} \qquad (5)$$

where $P(r_j)$ and $P(\beta)$ are the prior probability distributions for the scale radius and the isotropy parameter. The scale radius and the isotropy parameter are assumed to be statistically independent. For readers who dislike the Bayesian language, the expression is identical to the likelihood function for the parameters with the prior probabilities representing changes of variables. The marginal probability distribution for one variable is found by projecting the two-dimensional probability distribution onto the variable, and the marginal distributions for the isotropy and scale radius are $P(r_j|v_{ri},I) = \int d\beta P(r_j,\beta|v_{ri},I)$ and $P(\beta|v_{ri},I) = \int dr_j P(r_j,\beta|v_{ri},I)$. From here on we drop the denominator used to normalize the Bayesian probability expressions and use a proportionality sign. All probability distributions have an implied denominator that normalizes the total probability to equal one.

We assume that the prior probability distribution for the scale radius is logarithmic, $P(r_j) = 1/r_j$ (corresponding to a prior on the galactic mass of $P(M) = 1/M$ as used by LT and ZI), so that we select no preferred scale for the radius, and we truncate the distribution at 10 kpc and 1 Mpc. A uniform prior for the isotropy $P(\beta) = 1$ is a bad choice in a Baysian analysis because of the asymmetric treatment of radial and circular orbits in the definition of $\beta = 1 - \sigma_\theta^2/\sigma_r^2$. There is "infinite space" for tangential orbits ($\beta < 0$) in the variable $\beta$ because its range is $[-\infty, 1]$. This means that the marginalization procedure combined with a uniform prior on $\beta$ will overweight tangential orbits. We instead use a "uniform energy" prior, $P(\beta) = (3 - 2\beta)^{-2}$, which asserts that the probability distribution of the ratio of the radial kinetic energy to the total kinetic energy is uniform. We limited the range for the isotropy to $-1 < \beta < 1$, and in this limited range we can compare a uniform isotropy prior to the energy prior.

## 2.1 Radial Velocities Only

Our first analysis uses only the satellite radial velocities. There are 25 satellites with galactocentric radii larger than 20 kpc and measured radial velocities. Of these, 19 are at radii larger than 30 kpc, and 15 are at radii larger than 50 kpc. Nine are dwarf spheroidals (all outside 50 kpc) and 16 are globular clusters. Table 1 summarizes the data. The primary systematic uncertainty in the data is the treatment of the Leo I dwarf spheroidal, which is both distant (230 kpc) and rapidly moving ($v_r = 177$ km s$^{-1}$). We used a limit of 20 kpc because the fast moving globular clusters NGC 6715, Pal 13, and NGC 5824 are important constraints on small values for the Jaffe radius. In general we can neglect the velocity errors of $\lesssim 5$ km s$^{-1}$ on the radial velocities given the other uncertainties in the problem. Significant distance revisions such as that for Pal 15 (Seitzer & Carney 1990) are a far more important source of errors than the statistical errors in measuring radial velocities.

To convert the measured radial velocities into galactocentric radial velocities we used the standard galactic radius, rotation rate, and local peculiar velocities used by LT and ZI ($R_0 = 8.5$ kpc, $\Theta_0 = 220$ km s$^{-1}$, (u,v,w)=(-9,12,7) km s$^{-1}$). Following LT, we correct the radial velocity for the contamination by the tangential component introduced by the finite distance of the sun from the galactic center by setting the galactocentric radial velocity to be $v_r = v_{r\odot}/(1 - \beta \sin^2 \alpha)^{1/2}$ where $v_{r\odot}$ is the observed radial velocity in the LSR, and $\alpha$ is the angle between the vectors from the satellite to the sun and the satellite to the galactic center. The correction is small for the range of $-1 < \beta < 1$ even for satellites with galactocentric radii of 20 kpc. The energy condition for a non-zero probability $P(v_r|\beta, r_j)$ requires that $v_c^2 \ln(1 + r_j/r) - v_r^2/2 > 0$, and the resulting limits on $r_j$ are shown in Table 1.

Because of the powerful effects of Leo I, it is useful and traditional to compare the results with Leo I to the results without Leo I. Figure 1 shows the likelihood function $P(r_j, \beta|v_{ri}, I)$ and the marginal distributions for the isotropy and Jaffe radius for the model with $r_s = 250$ kpc, and Table 2 summarizes the results for all the different models of the radial distribution of satellites and the two different isotropy priors. The results with and without Leo I are disjoint. If we exclude Leo I the orbital distribution is predominantly radial with a median value of $\beta = 0.53$ and a 90% confidence range of $-0.57 < \beta < 0.94$.[2] The halo is smaller than the orbit of the LMC with a median Jaffe radius of $r_j = 44$ kpc and a 90% confidence range of $29\,\text{kpc} < r_j < 255\,\text{kpc}$. If we include Leo I the orbital distribution is predominantly isotropic with a median value of $\beta = -0.20$ and a 90% confidence range of $-0.90 < \beta < 0.48$. The halo is large, with a median Jaffe radius of $r_j = 334$ kpc and a 90% confidence range of $129\,\text{kpc} < r_j < 883\,\text{kpc}$. Here the upper limit on the halo size and the lower limit on the isotropy are controlled by the limits in the prior distributions at $r_j = 1$ Mpc and $\beta = -1$. If we use the uniform prior for $\beta$ instead of the energy prior, the orbits become more tangential and the halo radii increase. The results are insensitive to the assumed cutoff in the satellite density distribution $r_s$. If we try models with $r_s = 100$ kpc and $r_s = 500$ kpc we find that the median value of $\beta$ changes by $\sim 0.1$ and the median value of $r_j$ changes by $\sim 5\%$ where smaller values of $r_s$ increase $\beta$ and $r_j$. Models where $r_s = r_j$ and the satellites have the same distribution as the mass, give slightly smaller halos with Leo I, and slightly larger halos without Leo I. Models where $r_s = r_t$ give slightly lower median halo sizes.

---

[2] The 90% confidence interval is defined by the range of the variable that includes 90% of the probability, with 5% of the total outside both the maximum and minimum limits.

figures/fig1.ps

Fig. 1.—Likelihood function and marginal distributions for the Milky Way satellite radial velocity data. The solid lines exclude Leo I, and the dashed lines include Leo I. The likelihood contours are shown at 32%, 10%, 4.5%, and 1% of the peak. The projected Bayesian marginal distributions for $\beta$ and $r_j$ are shown below and right of the likelihood contour graph respectively. The median (squares) and 90% (triangles) probability intervals are marked on the marginal distributions. The probability is on a relative scale.

As we predicted in the introduction and was found by KL, the distribution of the line of sight velocities can crudely estimate the orbital isotropy. In particular the orbital distribution cannot be radial if we include Leo I, but it is nearly radial if we exclude Leo I. This (somewhat) counterintuitive result arises because none of the other outer satellites have appreciable radial velocities. Table I shows that the strongest constraints on $r_j$ besides Leo I are NGC 5694, NGC 7006, Pal 14, and Eridanus, and these only need halos extending to slightly beyond the edge of the disk. Without Leo I the best model is a small halo with quasi-radial velocities. If Leo I is bound, the average total velocities of the inner satellites need to be much higher than the average radial velocities for consistency, and this can only be achieved by giving the orbits a substantial tangential component. The larger the halo becomes, the greater the amount of kinetic energy that must be hidden in tangential motions. *If Leo I is bound, or nearly bound, the inner satellites must have large tangential velocities.*

### 2.2 The LMC, Sculptor, and Pal 3 Proper Motions

There are proper motion measurements for the LMC (Jones et al. 1994), Draco and Ursa Minor (Scholz & Irwin 1994), Sculptor (Schweitzer & Cudworth 1995), and Pal 3 (Cudworth 1993). Table 3 summarizes the proper motion data and its errors. The LMC, Sculptor, and Pal 3 all have total velocities comparable to $\Theta_0 = 220$ km s$^{-1}$ even though they are 50 kpc or more from the galactic center. The uncertainties in the LMC and Sculptor velocities are $\sim 50$ km s$^{-1}$, while the uncertainties in the Pal 3 velocity are $\sim 100$ km s$^{-1}$. Draco and Ursa Minor have both large velocity errors and extraordinarily high velocities (400-500 km s$^{-1}$), so we decided not to include them in the analysis.

Since the lower limit on the Jaffe radius depends exponentially on $(1/2)(v/v_c)^2$, 20% velocity errors are significant and we must account for the error bars on the proper motions. We include the errors by using the probability distribution for the radial and tangential velocities averaged over the measurement errors in the radial velocity, the proper motions, and the distance to the satellite. We assumed that the velocity and proper motion errors are Gaussian and have the stated one standard deviation errors. We included a 10% Gaussian distance error for the satellites other than the LMC and a 5% error for the LMC. Using these estimates of the distribution we generated 100 Monte Carlo realizations of each satellite by drawing the values of the radial velocity, the distance, and the proper motion from the Gaussian distributions. Each satellite was included in the calculation using the ensemble average of the realizations.[3]

figures/fig2.ps

Fig. 2.–Likelihood function and marginal distributions for the Milky Way satellite radial velocity data combined with the LMC, Sculptor, and Pal 3 proper motions. The solid lines exclude Leo I, and the dashed lines include Leo I. The likelihood contours are shown at 32%, 10%, 4.5%, and 1% of the peak. The projected Bayesian marginal distributions for $\beta$ and $r_j$ are shown below and right of the likelihood contour graph respectively. The median (squares) and 90% (triangles) probability intervals are marked on the marginal distributions.

Figure 2 shows the likelihood contours and the marginal probability distributions for $\beta$ and $r_j$ when $r_s = 250$ kpc, and the results for the other satellite density profiles are summarized in Table 4. The likelihood function with Leo I included is little changed, but the likelihood function without Leo I shows a marked shift of the distribution to larger Jaffe radii and more isotropic distribution functions. With the addition of the proper motion data the two distributions are not in complete agreement, but they no longer disjoint: the 90% confidence region without Leo I overlaps the one standard deviation region with Leo I. Notice that the absolute lower bound on the Jaffe radius does not increase because of the broad error bars on the proper motions – the minimum permissible escape velocity is still being set by the high radial velocity satellites. The median Jaffe radius in the models without Leo I increases to $r_j \simeq 89$ kpc in the standard model. Smaller satellite scale radii or the uniform isotropy prior increase the median to $r_j \simeq 120$ kpc. Where the radial velocities of these satellites allowed halos so small that they were inconsistent with the rotation curve of the disk, the tangential velocities inferred from the proper motions force the Jaffe radius to be larger than the disk.

## 3 ESCAPE VELOCITIES, TIDAL TRUNCATION, AND LOCAL GROUP TIMING

In §2 we examined the Milky Way satellites and explored the effects of isotropy, proper motions, and Leo I on estimates of the Jaffe radius for a fixed circular velocity of 220 km s$^{-1}$. In this section we explore fitting the circular velocity $v_c$ and adding additional physical and dynamical constraints to the models. To reduce the numbers of permutations of the satellite data, we adopt a standard model including the LMC, Sculptor, and Pal 3 proper motion data either with or without Leo I and a standard satellite scale radius of $r_s = 250$ kpc. Smaller cutoff radii modestly increase the satellite mass estimates, and larger cutoff radii modestly decrease the mass estimates (see Tables 2 and 4).

LT and ZI fit models with $v_c$ as a parameter in the limit that $r_j \to \infty$. They found median circular velocities ranging from 89 km s$^{-1}$ for radial orbits without Leo I, to 170 km s$^{-1}$ for tangential orbits with Leo I. These models are inconsistent with the known rotation curve of the Galaxy near $2R_0$. The low velocities should neither be interpreted as implying a low total mass for the Galaxy nor as implying a small circular velocity outside the disk. Let the satellite with velocity (radial or total) $v$ and radius $r$ set the strongest constraint on the model, then the threshold for $v_c$ and $r_j$ is set by the condition that the escape velocity at radius $r$ is larger than $v$, or $v^2 < v_e^2 = 2v_c^2 \ln(1 + r/r_j)$. As a function of $v_c$ this condition

---

[3] A similar calculation for the radial velocity analysis showed so little effect on the conclusions that we did not include it.

sets a lower bound on the total mass of the Galaxy

$$M_T(v_c) > M_{Tmin}(v_c) = \frac{v_c^2 r}{G} \left[ \exp\left(\frac{1}{2}\frac{v^2}{v_c^2}\right) - 1 \right]. \quad (6)$$

The smallest mass solution is found for $v_c \gg v$ where $M_{Tmin}(v_c) \to v^2 r/G$ while the highest mass solution is found for $v_c \ll v$ where $M_{Tmin}(v_c) \to (v_c^2 r/G)\exp(v^2/2v_c^2)$. Although a solution with $v_c \simeq 170$ km s$^{-1}$ implies a small mass interior to the LMC, it also requires a very extended halo. *Because the limits on $v_c$ and $r_j$ are controlled by the escape velocity, the models cannot be truncated in radius without also increasing the velocity estimate.* All the LT, ZI, and KL models either have infinite mass (slowly decreasing power law density density profiles) or are inconsistent with the rotation curve of the disk (many of the power laws as well as the point mass models). A major advantage of the Jaffe model over these scale free models is that it has a scale length that allows it to have finite mass and a flat rotation curve in the inner regions.

In this section we build a consistent model of the halo starting with the constraints from the satellites in §3.1. In §3.2 we add the constraints from the limits on the local escape velocity of stars. In §3.3 we add the constraints from the rotation curve of the disk, largely to eliminate solutions with rotation velocities near $2R_0$ that are too high. In §3.4 we consider the effects of "tidally" truncating the Galaxy halo at the classical Roche limit of $r_T \simeq 325$ kpc set by the separation and mass ratio of the Galaxy and M31. Mass that is bound to the Galaxy must (roughly) be confined to the region inside $r_T$, although the halo of the Galaxy may merge into a larger halo for the Local Group. Finally, in §3.5 we examine how the Local Group timing argument suppresses the high mass solutions. A complete summary of the models and their permutations is given in Table 5.

### 3.1 Satellite Orbits

Figure 3 shows the likelihood functions and the marginal distributions if we simply add $v_c$ to the analysis of §2 as a parameter with a prior probability distribution of $P(v_c) = 1/v_c^2$ over the range from 120 km s$^{-1}$ < $v_c$ < 320 km s$^{-1}$. LT and ZI used the same prior, and the combined prior $P(r_j)P(v_c) \propto 1/v_c^2 r_j \propto 1/M$ where $M$ is the total mass of the Galaxy. The results are shown using the proper motion data for the LMC, Sculptor, and Pal 3, and either with or without Leo I. The proper motion data and the upper limit on $r_j$ at 1 Mpc exclude the low velocity solutions found by LT and ZI. The median estimate is $v_c \simeq 210$ km s$^{-1}$ and the 90% confidence interval is 140 km s$^{-1}$ $\lesssim v_c \lesssim$ 300 km s$^{-1}$ independent of the treatment of Leo I. The marginal distribution for $\beta$ peaks near $\beta = 0.5$ in both cases, although the median isotropy is 0.24 without Leo I and 0.06 with Leo I. The marginal distribution for the Jaffe radius shows the only radical change caused by Leo I, with the median increasing from 138 kpc without Leo I to 436 with Leo I. For $v_c$ = 120 km s$^{-1}$, 220 km s$^{-1}$ and 320 km s$^{-1}$ the minimum masses of the Galaxy (excluding Leo I) are 8.4, 4.5, and 4.2 $\times 10^{11} M_\odot$. As we discussed in the introduction to §3, the low velocity solutions require the highest mass galaxies. The likelihood contours in the $v_c$-$r_j$ plane show few signs of closing in the limit of low velocities and large halos, and they are only beginning to close in the limit of high velocities and small halos. Thus the edge of the prior probability distribution $P(r_j)$ at $r_j = 1$ Mpc leads to the limits on low velocity dispersions. Some extra constraints are needed to force the contours to close away from the edge of the prior.

figures/sec4.fig1.ps

FIG. 3.—Constraints on $v_c$, $r_j$ and $\beta$ using satellite orbits. The two dimensional marginal probability distribution for $v_c$ and $r_j$, and the marginal probability distributions for $v_c$, $r_j$, and $\beta$ are shown. The solid (dashed) curves exclude (include) Leo I. The likelihood contours are shown at 32%, 10%, 4.5%, and 1% of the peak. The median (squares) and 90% (triangles) probability intervals are marked on the marginal distributions.

### 3.2 The Local Escape Velocity

Nearby high velocity stars estimate the local escape velocity $v_e(R_0)$. Leonard & Tremaine (1990) found a limit of 450 km s$^{-1}$ $\lesssim v_e(R_0) \lesssim$ 650 km s$^{-1}$ at 90% confidence using the Carney & Latham (1987, 1988) sample of high proper motion stars. The lower bound is determined by the highest velocity stars in the sample, and the upper bound is determined by how sharply the distribution function cuts off near the escape velocity and the stellar density distribution cuts off in radius. The lower bound on the escape velocity sets a lower bound on the Jaffe radius of 9.6 Mpc, 60 kpc, and 14 kpc for $v_c = 120$, 220 and 320 km s$^{-1}$. This limit rises more sharply for small circular velocities than the threshold for the satellites, so adding the escape velocities will strongly reduce the probability of low $v_c$ solutions.

The distribution function of the stars near the escape velocity must have the asymptotic form $f(\epsilon) \propto \epsilon^k$ where $\epsilon = (v_e^2 - v^2)/2 > 0$ (Leonard & Tremaine 1990). For example, if we expand equation (1) for the distribution function of the satellites in the limit of small $\epsilon$ and assuming a power law $r^{2\beta}\nu \propto r^{-\alpha}$ for the stars at large radii, we find that

$$f(\epsilon) \propto \epsilon^{\alpha-3/2} \left[ 1 + \frac{\alpha(1+\alpha)}{2\alpha - 1} \frac{\epsilon}{v_c^2} + \cdots \right], \tag{7}$$

and the expansion is valid if $\epsilon/v_c^2 = (v_e^2 - v^2)/2v_c^2 \ll 1$. If the distribution is isotropic, then the exponent $k = \alpha - 3/2$ where $\alpha$ is the power law slope of the density distribution. Tremaine (1987) and Jaffe (1987) argue that violent relaxation should produce $k = 3/2$, while Pritchet & van den Bergh (1994) found that the stellar halo of M31 drops of as $r^{-5}$ at 20 kpc from the center ($k \simeq 7/2$) and Sackett et al. (1994) found that the profile of NGC 5907 drops as $r^{-9/4}$ at 5 kpc above the disk ($k \simeq 3/4$). For the particular estimate of the isotropy in equation (1), the power law index of the density profile is $k + 3/2 - 2\beta$. Even if the distribution function is anisotropic, we can neglect the anisotropy if the stars are isotropically sampled on the sky and hence in viewing angles across the velocity ellipsoid. Leonard & Tremaine (1990) also found that analyzing only the radial velocities was as good an estimator of the escape velocity as analyzing the full space velocities because of the larger uncertainties in the tangential velocities from systematic and statistical errors in the proper motions and distances.

We approximate the distribution function near the escape velocity by $f(v) = (v_e^2 - v^2)^k$ including the $(v_e + v)^k$ term dropped by Leonard & Tremaine (1990) because it has a strong systematic variation with the exponent $k$. If we consider only the radial (meaning line of sight) velocities we must average over the unknown tangential velocities and then normalize the probability to include only stars exceeding a cutoff velocity $v_{cut}$ to get the likelihood for measuring line of sight $v_r$ between $v_e > v_r > v_{cut}$. If we ignore the dependence of the survey

figures/sec4.fig2.ps

Fig. 4.—Likelihood function and marginal distributions for the circular velocity $v_c$, the Jaffe radius $r_j$, and the stellar distribution function exponent $k_r$. The solid lines are the distributions using only the high velocity stars, and the dashed lines are the distributions using both the satellites and the high velocity stars. The likelihood contours are shown at 32%, 10%, 4.5%, and 1% of the peak. The median (squares) and 90% (triangles) probability intervals are marked on the marginal distributions.

volume on the tangential velocity, we find the Leonard & Tremaine (1990) result that

$$P(v_r|v_c, r_j, k) = \frac{(v_e^2 - v_r^2)^{k_{LT}}}{\int_{v_{cut}}^{v_e}(v_e^2 - v_r^2)^{k_{LT}} dv_r}, \qquad (8)$$

where $k_{LT} = k + 1$. This model ignores the selection only of stars exceeding a minimum proper motion in the Carney et al. (1987, 1994) samples, which makes the sample volume is a function of the tangential velocity. If the minimum proper motion is $\mu_{min}$, then stars are included out to radius $r = v_t/\mu_{min}$ from the sun, where $\mu_{min}$ is a function of the galactic coordinates of the star and the position angle of the proper motion vector. If this distance is always smaller than the depth limit caused by the flux limit for the stars (as is true for this sample), we find that

$$P(v_r|v_c, r_j, k) = \frac{(v_e^2 - v_r^2)^{k_r}}{\int_{v_{cut}}^{v_e}(v_e^2 - v_r^2)^{k_r} dv_r}, \qquad (9)$$

where $k_r = k + 5/2$ if $r_f \mu_{min} \gtrsim v_e$ where $r_f$ is the maximum distance set by the flux limit of the survey. The extra factor of $(v_e^2 - v_r^2)^{3/2}$ in $k_r$ compared to $k_{LT}$ arises because when $v_r \sim v_e$ the tangential velocity becomes small, $v_t = (v^2 - v_r^2)^{1/2} < (v_e^2 - v_r^2)^{1/2}$ so the volume in which the star has a large enough proper motion to be included in the sample shrinks as $r^3 \propto (v_t/\mu_{min})^3 \propto (v_e^2 - v_r^2)^{3/2}$.

We use the radial velocity data from the Carney et al. (1994) survey of proper motion stars. The sample has 31 stars with line of sight velocities in the LSR larger than 250 km s$^{-1}$, and 10 with velocities larger than 300 km s$^{-1}$. The Leonard & Tremaine (1990) analysis was based on 15 stars with line of sight velocities greater than 250 km s$^{-1}$. If we apply the Leonard & Tremaine (1990) model for the line of sight velocity ($k_{LT} = k + 1$, eqn. 8) to the Carney et al. (1994) data with a uniform prior on the range $1 < k_{LT} < 3$ we find that for $v_{cut} = 250$ km s$^{-1}$ the 90% confidence range for the escape velocity is 441 km s$^{-1} < v_e <$ 521 km s$^{-1}$, and for $v_{cut} = 300$ km s$^{-1}$ we find 449 km s$^{-1} < v_e < 626$ km s$^{-1}$. If we use our model for the line of sight velocity $k_r = k + 5/2$ and analyze the same data with a uniform prior $3 < k_r < 5$ (again bracketing the violent relaxation value of $k = 3/2$) we find that for $v_{cut} = 250$ km s$^{-1}$ the range is 466 km s$^{-1} < v_e < 579$ km s$^{-1}$ and for $v_{cut} = 300$ km s$^{-1}$ the range is 489 km s$^{-1} < v_e < 730$ km s$^{-1}$. The three largest space velocities in the sample are 610, 625, and 673 km s$^{-1}$ all for stars about 500 pc from the sun with proper motions near 0.″22 yr$^{-1}$. These stars are near the sample thresholds so the velocity uncertainties are appreciable ($\sim 50$ km s$^{-1}$), but it appears that the model including the proper motion volume limit (eqn. 9) is probably the more accurate and we use $k_r = k + 5/2$ from here on.[4]

For the full analysis we used a constant prior probability for the exponent $P(k_r)$ over the range $0.1 < k_r = k + 5/2 < 10$. This range for $k_r$ spans not only the range of uncertainty for the stellar distribution, but also the differences between $k_r$ and $k_{LT}$. We used $v_{cut} = 300$ km s$^{-1}$, choosing larger statistical errors from the small number of stars over

---

[4]We also redid the analysis keeping the second order terms predicted by equation (7). For our model of the line of sight velocity, and $v_{cut} = 300$ km s$^{-1}$ the 90% confidence region is 456 km s$^{-1} < v_e < 670$ km s$^{-1}$.

the larger systematic errors from using $v_{cut} = 250$ km s$^{-1}$. Bayes theorem then tell us that $P(v_c, r_j, k|v_{ri}, I) \propto P(v_c)P(r_j)P(k)\Pi_{i=1}^{N}P(v_{ri}|v_c, r_j, k)$ where there are $N$ stars in the sample with line-of-sight velocities $v_{ri}$ above $v_{cut}$. Figure 4 shows the likelihood function for fitting the high velocity stars as well as the marginal distributions for $v_c$, $r_j$, and $k_r$ using only the escape velocity

As expected, the local escape velocity suppresses the low velocity solutions by forcing the Jaffe radius to be significantly larger than 1 Mpc. The satellites are slightly stronger constraints on high velocities with small Jaffe radii and the limits are similar when $v_c \simeq \Theta_0$. The maximum size of the halo remains ill-determined because of the systematic uncertainties in $k$ and $\beta$, but the two independent types of data agree on the permitted regions of parameter space. The combined data gives a median $v_c = 237$ km s$^{-1}$ with a 90% confidence limit of 171 km s$^{-1} \lesssim v_c \lesssim 309$ km s$^{-1}$ where the lower bound is a combination of the constraints imposed by the data and the edge of the Jaffe radius prior, and the upper bound is set by the edge of the velocity prior.

As was found by Leonard & Tremaine (1990), the stellar velocity data alone cannot estimate the value of the exponent $k_r$. The marginal probability distribution is a slowly declining function of $k_r$ and even that decline is produced by the boundaries of the $r_j$ and $v_c$ prior distributions. When we combine the local escape velocity and the satellite constraints, we find that the marginal distribution of $k_r$ begins to be constrained. The median value of $k_r$ drops from 4.8 to 4.0, and the 90% confidence interval shrinks from $1.0 < k_r < 9.4$ to $0.8 < k_r < 8.9$. While we would be hard pressed to call this a measurement of $k_r$, it is nonetheless reassuring that as we try to assemble a globally consistent model of the halo the unknown parameters in the individual pieces show signs of converging to reasonable values.

### 3.3 The Rotation Curve of the Disk

The combined analysis of the satellites and the escape velocity limits the existence of physically reasonable low velocity models, but only weakly constrains the existence of high velocity, small Jaffe radius models (see Figure 4). Although the likelihood contours are beginning to close near $v_c = 320$ km s$^{-1}$, the upper bound on $v_c$ is strongly influenced by the prior. These solutions are, however, inconsistent with the rotation curve of the disk inside $2R_0$. The solution with $v_c = 320$ km s$^{-1}$ and $r_j = 25$ kpc implies a circular velocity at $R_0$ near 275 km s$^{-1}$ – much larger than can be covered up by appealing to the differences between flattened disks and spherical halos.

From the analysis by Fich et al. (1989) and the review by Fich & Tremaine (1991) we adopt estimates for the circular velocity at $R_0$ of $\Theta(R_0) = 220 \pm 22$ km s$^{-1}$ and for the circular velocity at $2R_0$ of $\Theta(2R_0) = 220 \pm 33$ km s$^{-1}$. The errors are somewhat overestimated to compensate for the differences between spherical models and disks. Remember that our mass model is meant to be the total mass of the Galaxy, not just the halo. We add the disk constraints using two Gaussian error distributions in the likelihood function

$$P(\Theta(2R_0)|r_j, v_c) \propto \exp\left[-\frac{1}{2}\left(\frac{\Theta(2R_0) - \Theta_{2R_0}}{\Delta\Theta_{2R_0}}\right)^2\right] \qquad (10)$$

figures/sec4.fig3.ps

FIG. 5.—Constraints on $v_c$, $r_j$ and $\beta$ using the satellite orbits, the stellar escape velocity and the disk constraints. The two dimensional marginal probability distribution for $v_c$ and $r_j$, and the marginal probability distributions for $v_c$, $r_j$, and $k_r$ are shown. The solid (dashed) curves exclude (include) Leo I. The likelihood contours are shown at 32%, 10%, 4.5%, and 1% of the peak. The median (squares) and 90% (triangles) probability intervals are marked on the marginal distributions.

$$P(\Theta(R_0)|r_j, v_c) \propto \exp\left[-\frac{1}{2}\left(\frac{\Theta(R_0) - \Theta_{R_0}}{\Delta\Theta_{R_0}}\right)^2\right] \quad (11)$$

where $\Theta_{2R_0} = \Theta_{R_0} = 220$ km s$^{-1}$, $\Delta\Theta_{2R_0} = 33$ km s$^{-1}$, and $\Delta\Theta_{R_0} = 22$ km s$^{-1}$.

Figure 5 shows the probability distributions using the satellites, the high velocity stars and the disk constraints. The primary effect of the disk constraints is to eliminate almost all the high velocity solutions, although there is a small, low probability tail for the models without Leo I. Models can only have low circular velocities in the halo by having a small Jaffe radius. The disk, satellite, and stellar velocity data work together to reduce the likelihood of low circular velocities, and the lower limit on $v_c$ is smaller than it would be based on the disk constraints alone, while they oppose each other on the high velocity end and the upper limit on $v_c$ is higher than it would be based on the disk constraint alone. Leo I has little effect on estimates of $v_c$, but it forces much larger Jaffe radii, larger stellar distribution function exponents, and more tangential satellite orbits.

### 3.4 Tidal Truncation & Cosmological Limits

The models in §3.3 are effective at constraining the minimum mass of the Galaxy, but ineffective at constraining the maximum mass because the lower limits are associated with sharp thresholds on the escape velocity, while the upper limits only close logarithmically in $r_j$. We saw in §3.1 that the 1 Mpc upper bound on $r_j$ restricts low $v_c$ solutions that require enormous halos. The 1 Mpc limit is, however, generous, since M31 is twice as massive as the Galaxy and only 730 kpc away. If the mass of the Local Group can be assigned to the two galaxies rather than to a larger group halo, then the outer edges of the individual halos must be truncated near the classical Roche limit for the Galaxy of $r_T \simeq 325$ kpc.

We impose a tidal limit by modifying the prior on $r_j$ to have a linear cutoff between $r_j = 275$ kpc and $r_j = 375$ kpc. This overestimates the size of a true tidally limited halo because half the mass of a Jaffe model lies outside $r_j$. The new limit will increase the lower limit on $v_c$ and eliminate the high mass solutions for the Galaxy. As a result the average orbital isotropy will become more radial and the average stellar distribution function exponent will decrease. Table 5 shows several examples of the parameter changes when the halo is truncated. These changes in the confidence limits should be regarded as qualitative because there is no quantitative theory for exactly how the halo of the Galaxy and the halo of the Local Group merge.

### 3.5 Local Group Timing

The last constraint we consider is the Local Group timing model of the Galaxy and M31 as a bound, binary system. We assume a fixed mass ratio of $x = M_{M31}/M_G = 2$, which is consistent both with the estimated ratio of luminosities and estimates from the circular velocities (eg. Peebles et al. 1989). The "classical" Local Group timing model of Kahn & Woltjer (1959) assumes that the orbits are radial and provides lower bounds on the mass. We add an uncertain ellipticity to the orbit in our analysis, following the approach of Einasto &

figures/sec4.fig4.ps

figures/sec4.fig4b.ps

Fig. 6.–Likelihood function and marginal distributions for the circular velocity $v_c$ and the Jaffe radius $r_j$, as well as the marginal distributions for $\beta$, $k_r$, and $e$. The dashed lines show the results using the disk, satellite, and high velocity star data. The likelihood contours are shown at 32%, 10%, 4.5%, and 1% of the peak. The median (squares) and 90% (triangles) probability intervals are marked on the marginal distributions.

Lynden-Bell (1982). Later studies by Peebles et al. (1989) and Raychaudhury & Lynden-Bell (1989) examined the effects of the collapse phase of evolution and the perturbations from other Local Group members, and concluded that these complications do not significantly affect the mass estimates for the Local Group from the timing argument.

We adopt $r_{M31} = 725 \pm 30$ kpc (eg. van den Bergh 1991) as the distance to M31, and a radial velocity (for $\Theta_0 = 220$ km s$^{-1}$) of $v_r = -123$ km s$^{-1}$ (Tully 1988). We assign an uncertainty of 30 km s$^{-1}$ to $v_r$ to represent uncertainties in $\Theta_0$ and $r_{M31}$ rather than measurement error. We assign M31 a tangential velocity $v_t = 60 \pm 30$ km s$^{-1}$ following the arguments of Einasto & Lynden-Bell (1982). Given the orbital ellipticity $e$ and phase $\eta$ we can compute the velocities and time since the first pericenter ($\eta = 0$) to be

$$v_r = v_c(1+x)^{1/2}\left(\frac{r_j}{r_{M31}}\right)^{1/2}\frac{e\sin\eta}{(1-e\cos\eta)^{1/2}} \quad (12)$$

$$v_t = v_c(1+x)^{1/2}\left(\frac{r_j}{r_{M31}}\right)^{1/2}\left(\frac{1-e^2}{1-e\cos\eta}\right)^{1/2} \quad (13)$$

$$t = \left(\frac{r_{M31}^3}{(1+x)v_c^2 r_j}\right)^{1/2}(\eta - e\sin\eta)(1-e\cos\eta)^{-3/2}. \quad (14)$$

We also know from cosmology (eg. Peebles 1980) that the current age of the universe is

$$t = H_0^{-1}f(\Omega) = H_0^{-1}(1-\Omega)^{-1}\left[1 - \frac{\Omega}{2(1-\Omega)^{1/2}}\cosh^{-1}\left(\frac{2}{\Omega}-1\right)\right] \quad (15)$$

if the cosmological constant $\lambda = 0$. The three measurements the model must fit are the two components of the velocity and the value of the Hubble constant, given the model parameters $\eta$, $e$, $\Omega$, $r_j$ and $v_c$. We use the standard priors for $r_j$ and $v_c$, and we assume uniform priors for $\eta$, $e$, and $\Omega$ over the ranges $\pi < \eta < 2\pi$ (M31 on its first orbit), $1/2 < e < 1$, and $0 < \Omega < 1$. The likelihood function $P(v_r, v_t, H_0|\eta, e, \Omega, r_j, v_c)$ consists of three Gaussian error functions for fitting $v_r = -123 \pm 30$ km s$^{-1}$, $v_t = 60 \pm 30$ km s$^{-1}$, and $H_0 = (74 \pm 20)$ km s$^{-1}$ Mpc$^{-1}$. We chose the Baysian estimate for the value of $H_0$ found by Press et al. (1995) but lacked the courage to use the formal error estimate of $\pm 4$ km s$^{-1}$ Mpc$^{-1}$ – the error bars were broadened to satisfy all camps. By Bayes theorem, the probability of the parameters given the data is then

$$P(\eta, e, \Omega, r_j, v_c|v_r, v_t, H_0, I) \propto P(\eta)P(e)P(\Omega)P(r_j)P(v_c)P(v_r, v_t, H_0|\eta, e, \Omega, r_j, v_c). \quad (16)$$

We regard $\eta$ and $\Omega$ as nuisance variables. The phase $\eta$ has a narrow uncertainty given the age and halo parameters, and the data cannot constrain $\Omega$ unless we narrowly restrict the Hubble constant, so we automatically marginalized the distribution over $\eta$ and $\Omega$ to estimate $P(e, r_j, v_c|v_r, v_t, H_0, I)$.

Adding the timing model eliminates all the high mass models of the Galaxy with large Jaffe radii, and the solutions with and without Leo I are mutually consistent because the timing model eliminates the small Jaffe radius models in the models without Leo I. Figure 6 shows the marginal probability distributions for all five of the model variables.

The constraints are almost entirely determined by the disk constraints (effectively $v_c = 220 \pm 30$ km s$^{-1}$) and the Local Group timing constraint (effectively $r_j = (180 \pm 60)(\Theta_0/v_c)^2$ kpc). The solutions permitted by these two constraints lie in the center of the broad plateau in the likelihood for fitting the satellites and the high velocity stars near $v_c \simeq 220$ km s$^{-1}$ and $r_j \gtrsim 100$ kpc. The satellite and escape velocity data do not significantly affect the error bars of the model.

We took a solution near the best model for the satellites and the mass distribution ($\beta = 0$, $r_j = 180$ kpc, and $v_c = 220$ km s$^{-1}$) and made Monte Carlo models of the satellite orbits by randomly chosing the satellite tangential velocities using the distribution function constrained by the observed radial velocities. We integrated the orbits for long periods (10 to 50 Gyr) to randomize the orbital phases, and then looked at the velocities and radii to see if the data would resemble what we observe. Qualitatively the data is always similar. We see roughly the same numbers of galaxies with radial velocities above 100 km s$^{-1}$ and 200 km s$^{-1}$ and orbital radii below 50 and above 100 kpc. There are usually several galaxies with tangential velocities above 200 km s$^{-1}$ and they are usually found at radii between 40 and 80 kpc. If we compute the minimum Jaffe radius from the radial velocities (as done in Table 1) the largest value in ten trials was 88 kpc, but the median was about 45 kpc and the lowest was 25 kpc. If the largest value is rejected (like throwing out Leo I), then the median value was 31 kpc with a range from 18 to 75 kpc. In short, the pattern seen in Table 1 for the limits is typical of orbits drawn from a galaxy in which $r_j = 180$ kpc.

## 4 Conclusions

We examined how the structure of our Galaxy is constrained by the orbits of satellites, the local escape velocity of stars, and Local Group timing while constraining the model to be consistent with the known dynamics of the disk. We use a Jaffe (1983) model for the total mass distribution of the Galaxy, the simplest density distribution that is globally consistent with a flat rotation curve and finite total mass. The Jaffe model depends on two parameters, $v_c$ and $r_j$, where $v_c$ is the circular velocity of the flat rotation curve for radii inside the cutoff radius $r_j$. The total mass of the Jaffe model is $v_c^2 r_j/G$. Previous halo models (eg. LT, ZI, KL) used power laws or point masses that were either inconsistent with the rotation curve of the disk, had infinite mass, or were much more extended than the M31/Galaxy separation. The fundamental question we answer is whether a mass distribution consistent with the disk at $r \simeq 2R_0$ shows evidence for an edge or a significant drop from a flat rotation curve by $r \simeq 100$ kpc when we include all the major systematic errors in the statistical analysis.

The two systematic problems in analyzing the satellites of the Galaxy are the isotropy of the orbits and whether Leo I is bound. By combining a statistical approach that is sensitive to the isotropy with the proper motions of the LMC, Sculptor, and Pal 3, we largely resolve these two issues. The satellite orbits are near isotropic, and the most natural models have Leo I bound to the galaxy. Monte Carlo simulations of the satellites drawn from a galaxy model with isotropic orbits, $v_c = \Theta_0 = 220$ km s$^{-1}$ and $r_j = 180$ kpc closely reproduce the observed properties of the satellites. Adding more proper motions or increasing the accuracy of the proper motions will rapidly reduce the remaining uncertainties.

figures/sec5.fig1.ps

Fig. 7.—The radial distribution of mass implied by the models of §3. The solid (dashed) lines show the median and 90% confidence intervals for the mass as a function of radius without (with) Leo I given the constraints on the $r_j$-$v_c$ parameter space. The sequence stars at the top left with only the satellite constraints and then successively adds the escape velocity data, the disk constraints, and the Local Group timing model. The heavy solid line shows the mass distribution for a constant rotation velocity of $\Theta_0 = 220$ km s$^{-1}$.

When the circular velocity near the disk is a parameter of the fit, the satellites allow considerable uncertainty in the mass distribution (see Figure 7). The reason is that the satellite data primarily constrains the escape velocity at the radius of each satellite, with much weaker constraints on the enclosed mass. If the halo velocity is small ($v_c \ll \Theta_0$) the halo must be very extended, reaching $r_j \simeq 1$ Mpc for $v_c = 100$ km s$^{-1}$. These solutions cannot be truncated at smaller radii and they correspond to the highest (total) mass models for the galaxy. Only by avoiding the scale-free models of the earlier satellite studies can we can explicitly see the anti-correlation of the mass inside the LMC radius with the total mass of the galaxy.

Next we added the constraints found by fitting for the local escape velocity of high velocity stars from the Carney & Latham (1994) sample using a modified version of the techniques introduced by Leonard & Tremaine (1990). Like the satellite orbits, the local escape velocity depends on an unknown systematic parameter. For velocities near the escape velocity the distribution function must approach a power law $f(\epsilon) \propto \epsilon^k$ where in an isotropic model the density distribution of the stars at large radii is $\propto r^{-k-3/2}$. The stellar velocity data alone are insufficient to estimate $k$, but if we assume a uniform prior distribution for the range $3 < k_r < 5$ and select stars with line-of-sight velocities above 300 km s$^{-1}$ we find a 90% confidence estimate for the local escape velocity of 489 km s$^{-1} < v_e(R_0) < 730$ km s$^{-1}$ using the Carney & Latham (1994) sample of stars. The limits are consistent with the highest, but uncertain, space velocity of 673 km s$^{-1}$ found by Carney & Latham (1994). The stellar velocities limit low $v_c$ models of the halo more strongly than the satellites, although the two data sets are broadly consistent. Figure 7 shows the reduced uncertainties in the mass as a function of radius using the two constraints. The marginal probability for the exponent $k$ now has a clear peak (Figure 6), and its median value of $k_r = k + 5/2 = 4$ is at the value expected for a violently relaxed distribution function.

While the median velocity solutions for the satellites and the escape velocity are consistent with the rotation curve of the disk, the outlying solutions are inconsistent with the observed rotation velocity near $2R_0$. Note, however, that the primary problem is with models that have too much mass inside $2R_0$ ($v_c$ too large). When we impose the constraint that the models loosely fit the circular velocities of the disk at $R_0$ and $2R_0$, the velocity parameter $v_c$ is set by the properties of the disk. The satellite and escape velocity data only determine the minimum Jaffe radius and its probability distribution. As illustrated by Figures 5 and 7, the primary effects of the disk constraints are to lower the upper mass bound and the median mass leaving the lower mass bound little changed. The median mass inside 50 kpc is 4.3 (5.4) with a 90% confidence region of 3.3 to 6.1 (4.2 to 6.8) without (with) Leo I in units of $10^{11} M_\odot$. This includes the proper weighting of the unknown systematic variables (the satellite isotropy and the distribution function exponent) and the uncertainties in the rotation curve of the disk. A Galaxy with a flat rotation curve and $\Theta_0 = 220$ km s$^{-1}$ has a mass interior to 50 kpc of $5.1 \times 10^{11} M_\odot$, so the mass interior to the LMC can be no lower than 65% of the amount expected if we simply continued the inferred rotation curve of the disk. The models are also consistent with models of the LMC/SMC/Magellanic Stream system – Lin et al. (1995) estimate that the mass inside 100 kpc is $5.5 \times 10^{11} M_\odot$.

With these three constraints the rotation curve is essentially flat to the LMC radius. The fundamental uncertainty is that the upper bounds on $r_j$ are only logarithmically constrained so that the upper bound on the mass of the Galaxy is uncertain. M31 can constrain the high mass solutions either by "tidally truncating" the halo, or through the Local Group timing constraint on the Galaxy and M31 as a binary galaxy system. At some level M31 must limit the size of our halo, and the characteristic scale where the halo must terminate or merge with a lower density (but possibly higher total mass) Local Group halo must be roughly the classical Roche limit of 325 kpc. Purely baryonic models of the halo must also exclude these solutions because they imply that the total contribution of spiral galaxies to the cosmological density exceeds the upper bound of the nucleosynthesis constraints. The timing mass estimates requires $r_j \simeq (180 \pm 60)(\Theta_0/v_c)^2$ kpc if M31 is currently on its first orbit, and when we add the Local Group timing model, including the uncertainties in $H_0$, $\Omega$, and the ellipticity of the orbit, we find there is a common solution for all the halo, with $v_c \simeq 230 \pm 30$ km s$^{-1}$ and $r_j \simeq 180$ kpc with 100 kpc $\lesssim r_j \lesssim 300$ kpc. Both error estimates are at 90% confidence. In this solution the limits with and without Leo I are almost identical (as was noted by ZI). The primary effect of adding the Local Group timing model is to *lower* the average mass of the Galaxy by eliminating the solutions with large Jaffe radii. If we exclude Leo I, the Local Group timing model also raises the 90% confidence lower bound on the mass inside 50 kpc from 3.2 to 3.8 in units of $10^{11} M_\odot$. The uncertainties in the model are largely determined by the disk constraints and the Local Group timing model, although the peak likelihood for these two constraints nicely overlaps the peak likelihood for the satellite data and the escape velocity.

There is a higher mass solution if M31 is on its second orbit, but this requires a halo with $r_j \simeq 400$ kpc. Once M31 begins making multiple, complete orbits, the halo will be tidally limited, and the mass required in the two orbit Local Group timing model must be partially distributed in a large Local Group halo. There is a lower mass solution that requires that M31 and Leo I are unbound and that the scale radius lies in the narrow range 50 kpc $\lesssim r_j \lesssim 100$ kpc. In a spherical cosmological model, the velocity of an unbound radial orbit is between $(2/3)(r/t) < v_r < r/t$ where $r$ is the distance to the object and $t$ is the age of the universe. For Leo I $r/t = 23$ km s$^{-1}$ and for M31 $r/t = 73$ km s$^{-1}$ for $t = 10$ Gyr so finding a model in which the two are unbound yet have radial velocities of 177 km s$^{-1}$ and $-123$ km s$^{-1}$ respectively is challenging. We know of no members of the Local Group massive enough to alter the relative velocity of M31 and the Galaxy by the $\gtrsim 200$ km s$^{-1}$ difference between an unbound galaxy at that radius and a bound galaxy at that radius. Even the excessively contrived Local Group models of Byrd et al. (1994) with multiple collisions between Local Group members come nowhere near producing such large deviations.

While these calculations included many of the systematic and statistical uncertainties in the data and the models, they did not include all. The solar radius $R_0$ was fixed at 8.5 kpc to match the 1985 IAU models (Swings 1985), while $8.0 \pm 0.5$ kpc is more likely (Reid 1993). Shifts in $R_0$ mainly affect these calculations through changes in the value of $\Theta_0$. Although we allowed deviations from $\Theta_0 = 220$ km s$^{-1}$ in the mass model, they were not included in the estimated velocities of the stars, galaxies, and M31 relative to the LSR. The strong (usually exponential) dependence of the Jaffe radius limits on the space velocity probably makes this the most important source of error in the data that was not included in the model. The Galaxy model was purely spherical, leading to errors of up to about 20% by mass in the region with the disk. The errors in the halo region are smaller for any reasonable limit on the oblateness of the halo. The isotropy of the satellite orbits was fixed, and we neglected the (hard to quantify) constraint that the pericentric radii of the satellites must be large enough to avoid tidal disruption. The escape velocity analysis used only the line-of-sight velocities and a simplified model of the selection effects in the survey. The Carney & Latham (1994) sample is large enough to justify a more careful analysis. The Local Group timing model ignored the differences between the halos of the two galaxies and a more diffuse Local Group halo, as well as perturbations from other Local Group members and mass accretion.

Acknowledgements: This research was supported by an Alfred P. Sloan Foundation Fellowship. The author thanks T. Kolatt, A. Loeb, E. Maoz, W. Press, and P. Schechter for discussions, and A. Schweitzer and K. Cudworth for allowing the use of their Sculptor proper motion data prior to publication.

| Table 1: Milky Way Satellites | | | | | | |
|---|---|---|---|---|---|---|
| Satellite | $R$ | $v_r$ | $\ell$ | b | Type | $\min(r_j)$ |
| | kpc | km/s | deg | deg | | kpc |
| NGC 6715 | 22 | 163 | 6 | $-14$ | GC | 7 |
| Pal 13 | 24 | 140 | 87 | 43 | GC | 5 |
| NGC 5634 | 25 | $-76$ | 342 | 49 | GC | 2 |
| NGC 5824 | 25 | $-127$ | 333 | 22 | GC | 5 |
| Rup 106 | 27 | $-44$ | 301 | 12 | GC | 1 |
| Pal 8 | 28 | 28 | 14 | $-7$ | GC | 0 |
| NGC 5694 | 31 | $-238$ | 331 | 30 | GC | 25 |
| NGC 6229 | 32 | 22 | 74 | 40 | GC | 0 |
| Pal 15 | 36 | 148 | 19 | 24 | GC | 9 |
| NGC 7006 | 39 | $-166$ | 64 | $-19$ | GC | 13 |
| LMC+SMC | 51 | 61 | 284 | $-36$ | DS | 2 |
| Ursa Minor | 65 | $-88$ | 105 | 45 | DS | 5 |
| Pal 14 | 75 | 166 | 29 | 14 | GC | 25 |
| Draco | 75 | $-95$ | 86 | 35 | DS | 7 |
| Sculptor | 79 | 74 | 288 | $-83$ | DS | 5 |
| Sextans | 85 | 78 | 244 | 42 | DS | 6 |
| Eridanus | 85 | $-138$ | 218 | $-41$ | GC | 19 |
| Carina | 93 | 14 | 260 | $-22$ | DS | 0 |
| Pal 3 | 95 | $-59$ | 240 | 42 | GC | 4 |
| NGC 2419 | 98 | $-26$ | 180 | 25 | GC | 1 |
| Pal 4 | 108 | 54 | 202 | 72 | GC | 3 |
| AM$-$1 | 117 | $-42$ | 258 | $-48$ | GC | 2 |
| Fornax | 140 | $-34$ | 237 | $-66$ | GC | 2 |
| Leo II | 220 | 16 | 226 | 49 | DS | 1 |
| Leo I | 230 | 177 | 220 | 67 | DS | 88 |

NOTES: Most of the data was collected from ZI and KL, which contain the original references. The velocity of Rup 106 is from Da Costa, Armandroff, & Norris (1992).

| Table 1: Milky Way Radial Velocities |||||||||
| Case || With Leo I |||| Without Leo I ||||
| $r_s$ | $\beta$ prior | $r_j$ | 90% range | $\beta$ | 90% range | $r_j$ | 90% range | $\beta$ | 90% range |
| --- | --- | --- | --- | --- | --- | --- | --- | --- | --- |
| 50 | energy | 317 | [122,878] | 0.20 | [-0.71,0.84] | 56 | [32,358] | 0.69 | [-0.39,0.95] |
|  | uniform | 404 | [144,910] | -0.17 | [-0.90,0.60] | 98 | [37,660] | 0.16 | [-0.85,0.91] |
| 100 | energy | 331 | [125,884] | 0.00 | [-0.82,0.69] | 49 | [31,300] | 0.64 | [-0.46,0.95] |
|  | uniform | 402 | [146,911] | -0.30 | [-0.94,0.44] | 84 | [34,604] | 0.11 | [-0.87,0.88] |
| 250 | energy | 334 | [129,883] | -0.20 | [-0.90,0.48] | 44 | [29,255] | 0.53 | [-0.57,0.94] |
|  | uniform | 388 | [145,905] | -0.43 | [-0.97,0.27] | 72 | [31,534] | 0.03 | [-0.89,0.80] |
| 500 | energy | 324 | [127,876] | -0.29 | [-0.93,0.37] | 42 | [28,242] | 0.45 | [-0.63,0.89] |
|  | uniform | 371 | [141,897] | -0.49 | [-0.98,0.19] | 67 | [30,496] | -0.02 | [-0.91,0.74] |
| $= r_j$ | energy | 289 | [119,848] | -0.19 | [-0.91,0.61] | 54 | [34,226] | 0.68 | [-0.51,0.95] |
|  | uniform | 362 | [141,887] | -0.45 | [-0.97,0.32] | 90 | [38,506] | 0.06 | [-0.90,0.90] |
| $= r_t$ | energy | 306 | [124,865] | -0.35 | [-0.94,0.32] | 41 | [28,211] | 0.46 | [-0.63,0.90] |
|  | uniform | 350 | [137,886] | -0.53 | [-0.98,0.14] | 64 | [30,449] | -0.02 | [-0.91,0.75] |

NOTES:

| Table 3: Proper Motion Data ||||||||
| Satellite | $R$ | $v_\odot$ | $\mu_\alpha$ | $\mu_\delta$ | $v_r$ | $v_T$ | min($r_j$) |
|  | kpc | km/s | mas/yr | mas/yr | km/s | km/s | kpc |
| --- | --- | --- | --- | --- | --- | --- | --- |
| LMC | 51 | 61 | $1.37 \pm 0.28$ | $-0.18 \pm 0.27$ | $47 \pm 8$ | $216 \pm 50$ | 33 ( 9) |
| Sculptor | 79 | 107 | $0.72 \pm 0.22$ | $-0.06 \pm 0.25$ | $95 \pm 8$ | $199 \pm 58$ | 52 (13) |
| Pal 3 | 95 | -59 |  |  | $248 \pm 95$ | $252 \pm 85$ | 324 (49) |

NOTES: In the min($r_j$) column, the first values is the minimum allowed value of the Jaffe radius $r_j$ given the estimated velocities, and the second value (in parenthesis) is the minimum value for a two-standard deviation reduction in the estimated velocity. The sources of the data are Jones et al. (1994) for the LMC, Schweitzer & Cudworth (1995) for Sculptor, and Cudworth (1993) for Pal 3.

| Table 4: Milky Way Radial Velocities & LMC/Sculptor/Pal 3 Proper Motions |||||||||
|---|---|---|---|---|---|---|---|---|
| Case || With Leo I |||| Without Leo I ||||
| $r_s$ | $\beta$ prior | $r_j$ | 90% range | $\beta$ | 90% range | $r_j$ | 90% range | $\beta$ | 90% range |
| 50 | energy | 415 | [151,913] | -0.03 | [-0.81,0.55] | 154 | [52,759] | 0.03 | [-0.81,0.63] |
|  | uniform | 462 | [170,927] | -0.28 | [-0.93,0.39] | 205 | [62,819] | -0.28 | [-0.94,0.45] |
| 100 | energy | 420 | [155,913] | -0.16 | [-0.87,0.45] | 153 | [48,758] | -0.08 | [-0.87,0.58] |
|  | uniform | 462 | [172,926] | -0.39 | [-0.95,0.28] | 204 | [59,815] | -0.37 | [-0.96,0.38] |
| 250 | energy | 376 | [141,900] | -0.25 | [-0.91,0.38] | 89 | [37,594] | 0.01 | [-0.84,0.63] |
|  | uniform | 416 | [155,912] | -0.45 | [-0.97,0.21] | 127 | [44,709] | -0.32 | [-0.95,0.44] |
| 500 | energy | 357 | [136,892] | -0.32 | [-0.93,0.31] | 76 | [34,534] | 0.01 | [-0.85,0.62] |
|  | uniform | 396 | [149,907] | -0.50 | [-0.98,0.15] | 111 | [40,667] | -0.32 | [-0.96,0.44] |
| $= r_j$ | energy | 359 | [142,886] | -0.29 | [-0.93,0.39] | 119 | [50,606] | -0.05 | [-0.87,0.63] |
|  | uniform | 411 | [160,908] | -0.49 | [-0.98,0.20] | 162 | [60,719] | -0.37 | [-0.97,0.42] |
| $= r_t$ | energy | 335 | [132,879] | -0.36 | [-0.94,0.27] | 72 | [34,483] | 0.01 | [-0.86,0.62] |
|  | uniform | 375 | [144,897] | -0.53 | [-0.98,0.11] | 104 | [39,629] | -0.33 | [-0.96,0.44] |

NOTES:

| | | | | | Table 5: Halo Constraints | | | | | | | | |
|---|---|---|---|---|---|---|---|---|---|---|---|---|---|
| Sat | Esc | Dsk | Tid | M31 | $v_c$ ( km s$^{-1}$) | | $r_j$ (kpc) | | $\beta$ | | $k_r = k + 5/2$ | | $e$ |
| X | | | | | 210 | [138,302] | 138 | [31,775] | 0.24 | [-0.74,0.74] | | | |
| L | | | | | 208 | [142,292] | 436 | [120,925] | 0.06 | [-0.84,0.69] | | | |
| X | X | | | | 237 | [171,309] | 133 | [30,802] | -0.03 | [-0.84,0.57] | 4.0 | [0.8,8.9] | |
| | X | | | | 232 | [165,307] | 189 | [24,859] | | | 4.8 | [1.0,9.4] | |
| X | | X | | | 231 | [195,276] | 95 | [31,700] | 0.08 | [-0.77,0.64] | | | |
| | X | X | | | 226 | [195,266] | 258 | [40,871] | | | 4.9 | [1.2,9.2] | |
| X | X | X | | | 231 | [197,275] | 143 | [34,774] | -0.04 | [-0.83,0.57] | 3.9 | [0.9,8.8] | |
| L | X | X | | | 222 | [193,252] | 413 | [138,910] | -0.13 | [-0.84,0.48] | 3.6 | [2.4,9.4] | |
| X | | | X | | 223 | [150,306] | 95 | [29,288] | 0.25 | [-0.71,0.74] | | | |
| X | X | | X | | 252 | [189,312] | 87 | [28,281] | -0.08 | [-0.83,0.58] | 3.7 | [0.8,8.7] | |
| X | X | X | X | | 237 | [203,280] | 96 | [32,285] | 0.05 | [-0.76,0.60] | 3.2 | [0.8,7.9] | |
| | | | | X | 175 | [124,295] | 260 | [81,636] | | | | | 0.89 | [0.75, |
| X | | | | X | 195 | [131,296] | 199 | [71,534] | 0.00 | [-0.76,0.74] | | | 0.89 | [0.75, |
| | | X | | X | 225 | [191,260] | 161 | [83,284] | | | | | 0.89 | [0.75, |
| | X | | | X | 235 | [169,308] | 150 | [67,358] | | | 4.4 | [0.9,9.0] | 0.89 | [0.75, |
| X | X | | | X | 235 | [172,306] | 143 | [62,333] | -0.12 | [-0.86,0.52] | 4.2 | [0.9,8.9] | 0.89 | [0.75, |
| X | X | X | | X | 228 | [197,262] | 153 | [77,270] | -0.10 | [-0.81,0.48] | 4.1 | [1.4,8.2] | 0.89 | [0.75, |
| L | X | X | | X | 229 | [199,260] | 180 | [108,296] | 0.07 | [-0.67,0.57] | 4.7 | [1.8,8.6] | 0.89 | [0.74, |

NOTES: The first five columns show which constraints are used to limit the halo properties: Sat=Satellite data, Esc=local escape velocity, Dsk=disk rotation curve, Tid=tidally truncated halo at 325 kpc, M31=Local Group timing model. An X means the constraint is used, and an L in the Sat column means that Leo I was included in the calculation. The remaining columns show the median and 90% confidence regions for the parameters of the mass model, $v_c$ and $r_j$, the isotropy of the satellite orbits $\beta$, the power law index of the stellar distribution function $k_r = k + 5/2$ where $f(\epsilon) \propto \epsilon^k$, and the ellipticity of the M31/Galaxy binary orbit $e$.